\documentclass[prl,showpacs,twocolumn]{revtex4}
\usepackage{epsfig}
\font\tams                   = cmmib10  
\font\kleinhalbcurs          = cmmib10 scaled 833  
\font\bxf                    = cmbx10  
\font\sevenbf                = cmbx7  
\def\vec#1{{\textfont1=\tams\scriptfont1=\kleinhalbcurs  
\textfont0=\bxf\scriptfont0=\sevenbf  
\mathchoice {\hbox{$\displaystyle#1$}}{\hbox{$\textstyle#1$}}  
{\hbox{$\scriptstyle#1$}}{\hbox{$\scriptscriptstyle#1$}}}}  
 \def \vq{\vec{q}} 

 \def \oz {\overline z}

\def \b{\beta}

\def \cD{{\cal D}}

\def \be{\begin{equation}}  
\def \ee{\end{equation}}  
\def \bea{\begin{eqnarray}}  
\def \eea{\end{eqnarray}}  

\begin{document}   

\title{Universality in Glassy Low--Temperature Physics} 
\author{Reimer K\"uhn}
\affiliation{Institut f\"ur Theoretische Physik, Universit\"at Heidelberg, \\
Philosophenweg 19, 69120 Heidelberg, Germany}
\date{30. December 2001}
\begin{abstract}
We propose a microscopic translationally invariant glass model which exhibits
two level tunneling systems with a broad range of asymmetries and barrier
heights in its glassy phase. Their distribution is 
qualitatively different from what is commonly assumed in phenomenological
models, in that symmetric tunneling systems are systematically suppressed.
Still, the model exhibits the usual glassy low-temperature anomalies.
Universality is due to the collective origin of the glassy potential energy
landscape. We obtain a simple explanation also for the mysterious {\em
quantitative\/} universality expressed in the unusually narrow universal
glassy range of values for the internal friction plateau. \end{abstract}

\maketitle

The physics of glassy systems at low temperatures differs strikingly from
that of their crystalline counterparts \cite {ZePo71}. Differences are observed
in thermal properties, in transport phenomena, and in dielectric and acoustic
response, and are believed to be due to the  existence of tunneling
excitations with a broad range of energy splittings and relaxation times,
which are absent in crystals. The phenomenological standard tunneling model
(STM) \cite{And72}, and its generalization, the soft potential model (SPM)
\cite{Kar83}, which paraphrase this idea  quantitatively were for  many years
considered to provide a satisfactory rationalization of glassy low-temperature
physics.

Questions into the {\em origin of the universality\/} of these phenomena
\cite{YuLe88} have not yet found clearcut answers; see, however,
\cite{Bur+96}. In particular, the remarkable degree of {\em quantitative\/}
universality (e.g., the narrow range of values for sound-attenuation
in the 1 K regime) has remained a mystery so far. Also, relations between low-
and high-temperature phenomena could not possibly emerge from phenomenological 
descriptions. Moreover, a number of recent  experiments \cite{Nat98, Str98,
Cla+00} are difficult to reconcile with predictions of prevalent models.

In order to better understand the successes  and limitations of the
phenomenological approaches, we have recently proposed {\em microscopic\/} 
models for glassy low-temperature physics \cite{KuH97,KuU00} using 
heuristics taken from spin-glass theory. The purpose of the present letter is
to improve upon \cite{KuH97,KuU00} with respect to both modeling and scope of
analysis. We investigate a glass model described by the Hamiltonian $H =
\sum_i \frac{p_i^2}{2m} + U_{\rm int} (\{u_i\})$ with a {\em translationally
invariant\/} interaction energy of the form  \be   
U_{\rm int}=\frac{1}{4} \sum_{ij} J_{ij} (u_i-u_j)^2 + \frac{g}{2N} \sum_{ij}
(u_i-u_j)^4 \label{uint} 
\ee 
The $u_i$ designate deviations of particle coordinates from some 
reference positions, and (\ref{uint}) may be thought of as arising
from a Born von Karman expansion of the full interaction energy about these
positions -- taken  to be stationary but not necessarily stable.
As before \cite{KuH97,KuU00}, glassy properties are modeled by assuming
the expansion coefficients at the harmonic level to be  Gaussian random 
couplings with mean $\overline{J_{ij}}=J_0/N$ and variance
$\overline{(\delta J_{ij})^2}=J^2/N$. The non-random quartic
potential is for stabilization, and so requires $g>0$. Non-translationally
invariant versions \cite{KuH97} or models with partial translational 
invariance \cite{KuU00} have been introduced before.

Our main results are the following. (i) The model exhibits a glassy phase
at low temperatures. (ii) The potential energy landscape is represented
self-consistently through an ensemble of effective single-site potentials
which comprises both single-well potentials (SWPs) , and double-well
potentials (DWPs), the latter with a broad spectrum of barrier heights and
asymmetries. (iii) The distribution of parameters characterizing the
single-site potentials depends on the  collective state of the system, and
differs qualitatively from what is assumed in the phenomenological
models. (iv) The model exhibits  typical low temperature anomalies, e.g., of
the specific heat, and of acoustic or dielectric response. (v) Within the
model we can explain, in particular the quantitative universality of glassy
low-temperature anomalies expressed e.g. in the unusually narrow range of
values for the internal friction plateau. (vi) We also see some unusual
frequency dependence of internal friction curves similar to that recently
observed in vitreous silica \cite{Cla+00}.

We now turn to the analysis. Due to the translationally invariant interaction, 
the system contains global translations as zero modes, and the partition
function must be evaluated orthogonal to these modes, i.e. with the
constraint  $\sum_i u_i=0$.  Apart from this detail, the analysis follows
standard reasoning. The free energy is obtained  by averaging an $n$-fold
replicated partition sum, $-\b f(\b) = \lim_{N\to \infty,n\to 0}
(Nn)^{-1} \ln [Z_N^n]_{av}$, and can be expressed in terms of an
Edwards-Anderson matrix of two-replica overlaps $q_{ab}$. Although
four-replica overlaps must also be introduced during the calculation, the
final result does  not depend on them. The replica free energy reads     
\be
n\b f(\b) = \frac{(\b J)^2}{4} \sum_{ab} q_{ab}^2 - 3\b g \sum_a q_{aa}^2
- \ln \tilde Z_{n}
\ee
with $\tilde Z_{n}$ a single-site partition function corresponding
to the $n$-replica potential $U_{n} = \frac{1}{2} \sum_{a}
(J_0 +12 g\,q_{aa})\,u_a^2 -\frac{\b J^2}{2} \sum_{ab} q_{ab}
u_a u_b -\frac{\b J^2}{8}\big(\sum_a u_a^2\big)^2 + g \sum_a u_a^4$, 
that is, $\tilde Z_{n} = \int \prod_a d u_a \exp\{-\b U_{n}\}$. The order
parameters $q_{ab}$ must satisfy the fixed point equations $q_{ab} = \langle
u_a u_b \rangle$ with angle brackets denoting a Gibbs average w.r.t. the
potential $U_n$. As usual, to perform the $n\to 0$ limit, one
starts from parameterizations of the $q_{ab}$ matrix based on assumptions
concerning transformation properties of solutions under permutation of the
replica. 

We have looked at the replica symmetric (RS) solution  and a solution with
one step of replica symmetry breaking (1RSB). Both in RS and 1RSB  (and at all
higher levels of Parisi's RSB scheme) the system is described by an ensemble of
effective single-site potentials of the form  
\be
U_{\rm eff}(u)= d_1\,u +d_2\, u^2 + d_4 u^4\ ,
\label{ueff}
\ee
in which $d_4=g$ is the coupling constant of the quartic stabilizing
interaction, and $d_1$ and $d_2$ are randomly varying parameters. We shall
occasionally refer to $d_1$ and $d_2$ as to an effective local field  $h_{\rm
eff}$ and an effective harmonic coupling $k_{\rm eff}$ via $d_1 = -h_{\rm eff}$
and  $d_2=\frac{1}{2} k_{\rm eff}$. Their joint distribution is expressed in
terms of the order parameters of the system which are in turn obtained in terms
of Gibbs averages self-consistently evaluated over the $U_{\rm eff}$ ensemble
(\ref{ueff}). This ensemble of effective single-site potentials is a
representation of the glassy  potential energy landscape within a mean-field
description. Glassy low-temperature anomalies follow from considering the
effects of quantized excitations within this ensemble of local potentials.

In RS one assumes $q_{aa}=q_d$ and $q_{ab}=q$ for $a\ne b$. These must solve
the RS fixed point equations 
\be
q_d =\Big \langle \langle u^2\rangle \Big \rangle_{\oz,z}\ \ ,\qquad  
q =\Big \langle \langle u\rangle^2 \Big \rangle_{\oz,z}
\label{fpers}
\ee
in which inner brackets denote a Gibbs average w.r.t. the effective 
single-site potential (\ref{ueff})
with 
\bea
d_1&=&d_1^{\rm RS}=-J\sqrt q z\ ,\nonumber\\ 
d_2&=& d_2^{\rm RS}=\frac{1}{2}(J_0 + 12 g\, q_d -J^2 C +J~\oz)\ ,
\label{hkRS}
\eea
and $C=\b(q_d-q)$, and outer brackets an average  over the zero-mean,
unit-variance Gaussians $\oz$ and $z$ in $d_1$ and $d_2$. The ensemble of
single-site potentials thus comprises both  SWPs, and DWPs, the latter with
a broad spectrum of barrier heights. A glassy state is signaled by $q\ne 0$,
thus a non-degenerate distribution of asymmetries of single-site potentials.
In the present model the transition is continuous and the critical condition
is given by $1 = (\b_c J)^2 \big \langle \langle u^2\rangle^2 \big
\rangle_{\oz,z}$, which is just the de Almeida-Thouless condition for the
occurrence of a RSB instability. For models with symmetric coupling
distributions the transition temperature satisfies $T_c(J,g) = \frac{J^2}{g}
T_c(1,1)$ with $k_{\rm B}\,T_c(1,1)\simeq 0.133~E_*$ \cite{rem1}. In RS, $d_1$
and $d_2$ are Gaussian and uncorrelated. This would be qualitatively in line
with assumptions of the SPM, although there are quantitative differences. In
contrast to the STM, correlations are predicted between asymmetries and
tunneling matrix elements of tunneling systems in DWPs.

The RS solution is unstable and thus strictly not acceptable throughout the
glassy phase. For a non-translationally invariant model \cite{KuH97}, we
have shown RSB effects to be small for the specific heat. Here, we look in
greater detail also at the distribution of $h_{\rm eff}$ and $k_{\rm eff}$. We
shall find it to be {\em qualitatively\/} different from what has been
obtained in the RS approximation, and thus also from what is assumed in the
phenomenological models. 

In 1RSB, expected to exhibit the salient RSB effects, one assumes
$q_{aa}=q_d$, $q_{ab}=q_1$ for $1<|a-b|\le m$ and $q_{ab}=q_0$ for $|a-b|>m$.
The fixed point equations are now much more complicated, and we shall not
reproduce them here. Whereas $d_2$ has the same form as in RS, $d_2^{\rm 1RSB}
=\frac{1}{2}(J_0 + 12 g\, q_d -J^2 C +J~\oz)$, except that now
$C=\b(q_d-q_1)$, the local field is more complicated.  It is formulated in
terms of {\em two\/} variables $z_1$ and $z_0$   \be 
d_1=d_1^{\rm 1RSB}=  -J\sqrt{q_1-q_0}~z_1 - J\sqrt q_0~z_0 
\ee 
of which $z_0$ is a standard Gaussian, whereas $z_1$, while
deriving from a Gaussian, becomes correlated with $z_0$ and $\oz$ (thus
$d_2$) in an intricate way through $U_{\rm eff}$. In the low-temperature
limit, to which we restrict our attention here, the result is  
\be 
p(z_1|z_0,\oz) = \frac{\exp\{-z_1^2/2 -D~U_{\rm eff}(\hat u)\}}{\sqrt{2\pi}
\int \cD z_1 \exp\{-D~U_{\rm eff}(\hat u)\}}  
\ee
Here $D=\b m$, which has a finite $T\to 0$-limit, $\hat u = \hat
u(z_1,z_0,\oz)$ is the value of $u$ which {\em minimizes\/} $U_{\rm eff}(u)$ 
for given values of $z_0$, $z_1$,and $\oz$, and $\cD z_1$ is a Gaussian measure.
Fig. 1 displays the resulting distribution of $d_1$ {\em conditioned\/}
on $d_2$.

\begin{figure}[t] 
{\centering  
\epsfig{file=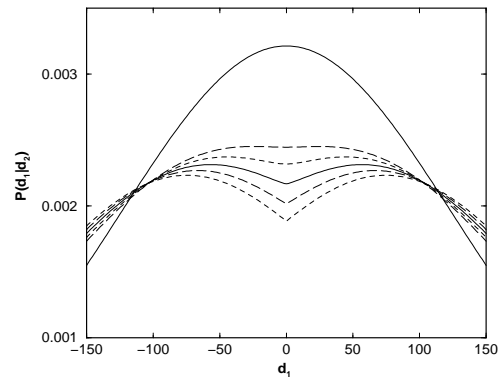,width=6.5cm, height=5cm} 
\par}  
\caption[]{Central part of $P(d_1|d_2)$, for $J=50$ and $g=1$. From top to
bottom we have RS (independent of $d_2$) and 1RSB with $d_2=10, 5, 0, -5$ and
$-10$, respectively}
\end{figure}

The RSB result is {\em qualitatively} different from the RS result and from
what is assumed in phenomenological models, in that small $d_1$ and
thus symmetric tunneling systems are systematically suppressed. In fact, in
the DWP region for negative $d_2$ and small $d_1$, the form is $P(d_1,d_2)\simeq
\pi_0(d_2) (1+\pi_1(d_2) |d_1|)$, so $P(d_1,d_2)$ is {\em non-analytic\/} at
$d_1=0$. Yet, due to the smallness of $\pi_1(d_2)$, RSB effects do not
have significant influence on functional forms of thermodynamic and response
functions at low temperatures. This was observed before for the specific heat
in a non-translationally invariant model \cite{KuH97}, and is confirmed here.

Glassy low-temperature anomalies follow from low energy excitations in the
ensemble of single-site problems $H_{\rm eff}= \frac{p^2}{2m} + U_{\rm
eff}(u)$. The energy of tunneling excitations in DWPs is determined by an
asymmetry $\Delta$ and a tunneling matrix element $\Delta_0$ as $E=
\sqrt{\Delta^2 + \Delta_0^2}$, which are in turn expressed through $d_1$ and
$d_2$ as $\Delta= E_* d_1\sqrt{2(|d_2|-1)}$ and $\Delta_0= E_* \sqrt 2
|d_2|^{3/2} \exp\{{2\over 3}(1-|d_2|^{3/2})\}$~\cite {RaB98}. Frequencies of
higher (quasi-harmonic) excitations in SWPs or DWPs are given by
the curvature of the potential in its minimum, $U_{\rm eff}''(u_{\rm min})=
m \omega^2$~\cite{rem1}. 

By averaging over $P(d_1,d_2)$ one obtains tunneling and vibrational density
of states and the specific heat as usual. Tunneling systems give rise to a
slightly {\em super-linear\/} specific heat at low temperatures, $C(T)\sim
T^{1+\varepsilon}$, with  $\varepsilon$ decreasing with $J$ (hence $T_c$), 
$\varepsilon \simeq$ 0.1, 0.05 and 0.01 for $J = 25$, 50, and 100 respectively,
whereas the strongly peaked vibrational density of state is the
origin of a Bose-peak in our model; see Fig. 2. RSB effects are small in
$C(T)$, the dominant effect being a  reduction of the tunneling
density of states as compared to RS. Except for the super-linearity at low $T$,
which can be traced down to the translational invariance of the interactions,
$C(T)$ is qualitatively as in \cite{KuH97}, both in RS and 1RSB.

\begin{figure}[t] 
{\centering  
\epsfig{file=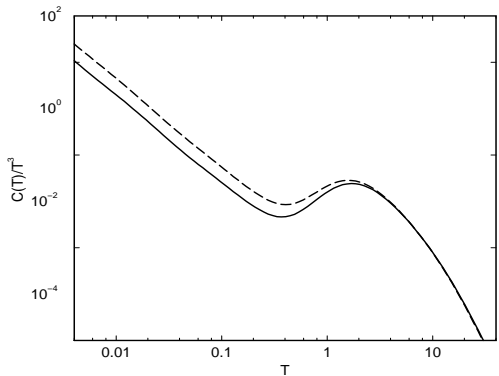,width=6.5cm, height=5cm} 
\par}  
\caption[]{Low $T$ specific heat for $J=50$ in RS (dashed) and 1RSB
(full).}   
\end{figure} 

It is interesting to compare with assumptions of the STM at the level of the
distribution  $P(\Delta,\lambda)$  where $\lambda =\ln(E_*/ \Delta_0)$. In
the STM $P^{\rm STM}(\Delta,\lambda)=P_0$ is assumed. We have  
\be
P(\Delta,\lambda)=\frac{P(d_1,d_2)}{E_* \sqrt{2(|d_2|-1)}
\left[\sqrt{|d_2|}-3/(2 |d_2|)\right]} 
\ee
for $d_2<-(3/2)^{2/3}$, with $d_1=d_1(\Delta,\lambda)$ and
$d_2=d_2(\Delta,\lambda)$ instead. Unlike in the STM or the SPM (i) the result
is not a constant, and (ii) symmetric tunneling systems are  systematically
suppressed. If one were to {\em fit\/} our results to an STM parameterization
one would associate $P_0 \simeq P(0,\lambda^*)$ for some $\lambda^* = {\cal
O}(1)\Leftrightarrow |d_2^*|= {\cal O}(10)$. Interestingly, for reasonable
energy scales (associated with tunneling of complexes like SiO$_2$ over atomic
distances, and giving $E_*$ between 1 and 5 K), we would estimate $P_0 =
10^{-7} \dots 10^{-6}~K^{-1}$/Atom for the system with $J=50$, which is the
right order of magnitude for typical glasses. Also, using the fact that the
low-$T$ limit of the fixed-point equations entails $q_d, q\propto J$, and 
$C\propto J^{-1}$ in RS, and similarly $q_d, q_1$ and $q_0\propto J$, 
$C\propto J^{-1}$, and $D\propto J^{-2}$ in 1RSB, we can combine this with
$T_c\propto J^2$ and the expression for $P(d_1,d_2)$ to predict the scaling
$P_0 \sim T_c^{-5/4}$ for large $T_c$ both in RS and 1RSB, our second result
relating low-$T$ properties with a property of the glass-transition in our
model. A detailed comparison with experiments is difficult, as a large change
in glass transition temperature requires changing composition of the glas,
and thus interactions in non-trivial ways. Yet the trend expressed by this
result appears to be correct.  

Next, we discuss dynamics due to a coupling between local degrees of freedom
$u$ and the strain field $e$ generated by Debye phonons, $H_{\rm SB}
= \gamma u e$, where $\gamma$ is a deformation potential and $e=\frac{1}{\sqrt
V}\sum_{\vq,s} \sqrt{\frac{\hbar}{2\rho\omega_{\vq,s}}}~iq (b_{\vq,s}-
b^\dagger_{\vq,s})$ the strain-field; the sum is over transversal and
longitudinal modes and the tensorial nature of the strain-field is neglected.
Quantities like internal friction are obtained by averaging the the dynamical
susceptibility over the ensemble of single-site problems,    \be
Q^{-1}  =  \frac{\gamma^2}{\rho v^2}\ \overline{\chi''_{uu}(\omega)}\ .
\label{frict}
\ee
\begin{figure}[t] 
{\centering  
\epsfig{file=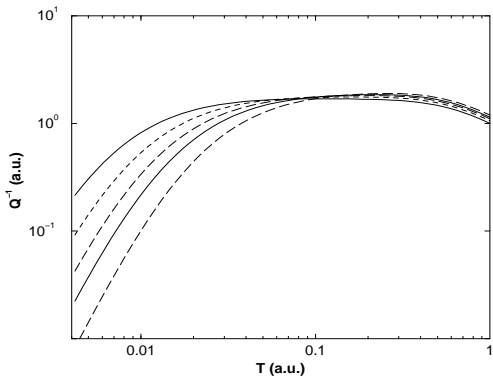,width=6.5cm, height=5cm} 
\par}  
\caption[]{Internal Friction as a function of $T$ on a double logarithmic
scale for various driving frequencies, $\omega=$ 0.33, 1.03, 2.52, 5.03 and 
14.0 kHz (top to bottom).} 
\end{figure} 
At low $T$, and for $\hbar\omega\ll k_{\rm B}T$, the dominant contribution is
the relaxational contribution of tunneling systems in DWPs within the ensemble
of single-site potentials. For these a two-level approximation for $H_{\rm
eff}$ is appropriate and (\ref{frict}) reduces to the well known expression
known for two-level tunneling systems, when the internal coordinate $u$ is
approximated by a two state variable in terms of a Pauli matrix, $u= u_0
\sigma_z$. Fig. 3 shows the internal friction of the present model for $J=50$
and driving frequencies as in \cite{Cla+00} computed in 1RSB (except for a
different global overall pre-factor, RS results are functionally hardly
distinguishable). Unlike in the STM and SPM, the dominant asymptotics is
$Q^{-1}\sim T^{3+\varepsilon}(1+ c_1 T)$ for $T \ll T^*$ and $Q^{-1}\simeq
\frac{\pi}{2} \frac{P_0\gamma^2}{\rho v^2}\left(\frac{\hbar\omega}{k_{\rm
B}T}\right)^{\varepsilon/2}(1+c_2 T)$ for $T\gg T^*$ --- the influence of the
$c_\alpha = {\cal O}(10^{-3})$ due to RSB effects being basically
undiscernable. The exponent $\varepsilon$ is that introduced before  to
quantify the super-linearity of the low-T specific heat, and $\varepsilon
\simeq 0.05$ for the $J$-value chosen. Note the following features: (i) there
is a slight frequency dependence $\omega^{\varepsilon/2}$  of the plateau
height, correlated with the low-$T$ specific heat exponent (the extra
$T^{-{\varepsilon/2}}$ factor is undiscernable for the small range of
temperatures considered) \cite{HoKu99}; (ii) the low $T$ asymptotics is has by
a temperature exponent slightly larger than 3. However, the crossover region
from plateau to low-temperature asymptotics is so large that effective
exponents in accessible temperature ranges are still {\em smaller\/} than 3,
the effect being stronger for lower frequencies. Both findings are well in
line with recent experiments  \cite{Cla+00}, though observed effective
exponents tend to show a somewhat stronger frequency dependence and be smaller
than ours.

Lastly, we turn to the universality issue. In \cite{Bur+96}, universality is
explained within a renormalization group approach as a collective effect due
to interactions {\em of\/} quantized low--energy excitations. Here it is
understood as a property of the interaction-generated glassy potential energy
landscape, thus as a collective effect {\em leading to\/} a particular
spectrum of quantized low--energy excitations. The mechanism is robust and may
therefore justly be expected to be insensitive to details. This holds in
particular for the tunneling density of states which is driven by a broad
distribution of asymmetries and barrier heights. Higher order excitations
depend to a larger extent on the shape of the stabilizing potential which is
not collectively modified to the same extent, and so a stronger material
dependence may be expected in this energy range; this fits well with the
observed partial loss of universality in the Bose peak region. Within the
present theory there is, beyond a rationalization of universality at the level
of exponents, also a rather simple explanation for the mysterious {\em
quantitative\/} universality of glassy low-temperature physics as expressed in
the unusually narrow range of values for the internal friction plateau. It
simply follows from combining the scaling $P_0\sim J^{-5/2}$ with the
plausible supposition $\gamma \sim J$ expressing the fact that deformation
potential and original interaction are of the same origin, and the elasticity
theory scaling $\rho v^2 \sim J$ for the sound velocity, which together entail
$\frac{P_0\gamma^2}{\rho v^2}\sim J^{-3/2} \sim T_c^{-3/4}$. This weak
parameter dependence due to cancellations implies, e.g., that the internal
friction plateau is changed by only a factor 8 when interaction energies are
changed such such as to increase $T_c$ from 100 to 1600 K!

In summary, we have proposed and analyzed a translationally invariant glass
model. It exhibits typical glassy low-temperature anomalies of specific heat
and acoustic attenuation. The super-linearity of the low-$T$ specific heat
is linked with the non-degenerate barrier-height distribution which in turn
can be shown to follow solely from translational invariance (a feature that
was absent in our original proposal \cite{KuH97}). Within the model we can
correlate low- and high-temperature properties. Universality is understood as
consequence of collective effects, and we also have a simple explanation of
the mysterious quantitative universality of internal friction data.  

This project profited a lot from a workshop on the low-temperature physics of
glasses held at the MPIPKS-Dresden. Helpful discussions with C. Enss, U. 
Horstmann, H. Horner,  S. Hunklinger, C. Picus and M. Thesen are also   
gratefully acknowledged.  


\begin{thebibliography}{99}

\bibitem {ZePo71}
R.C. Zeller and R.O. Pohl, Phys. Rev. B {\bf 4}, 2029 (1971)

\bibitem{And72} 
P.W. Anderson, B.I. Halperin, and C.M. Varma, Philos. Mag. {\bf 25}, 1 (1972);
W.A. Phillips, J. Low Temp. Phys. {\bf 7}, 351 (1972) 

\bibitem{Kar83} 
V.G. Karpov, M.I. Klinger, and F.N. Ignat'ev, Sovj. Phys. JETP {\bf 57}, 
439 (1983); U. Buchenau, Yu.M. Galperin, V.L. Gurevich, D.A. Parshin, M.A. Ramos, and
H.R. Schober, Phys. Rev. B{\bf 46}, 2798 (1992)

\bibitem {YuLe88}
C.C. Yu and A.J. Leggett, Comments Cond. Mat. Phys. {\bf 14}, 231
(1988)

\bibitem{Bur+96}
A.L. Burin and Yu. Kagan, JETP {\bf 82}, 159 (1996); Phys. Lett. A {\bf 215},
191 (1996)

\bibitem{Nat98}
D. Natelson, D. Rosenberg, and D.D. Osheroff, Phys. Rev. Lett. {\bf 80}, 4689
(1998)

\bibitem{Str98} 
P. Strehlow, C. Enss, and S. Hunklinger, Phys. Rev. Lett. {\bf 80}, 5361 (1998)

\bibitem{Cla+00} 
J. Classen, T. Burkert, C. Enss, and S. Hunklinger, Phys. Rev. Lett. {\bf 84},
2176 (2000) 

\bibitem{KuH97} 
R. K\"uhn and U. Horstmann, Phys. Rev. Lett. {\bf 78}, 4067 (1997)

\bibitem{KuU00} 
R. K\"uhn and J. Urmann, J. Phys. C {\bf 12}, 6395  (2000)

\bibitem{rem1} We set the energy scale $E_*$ by choosing $g=1$, thereby
relating $E_*$ via $E_*={\hbar^2/ m u_0^2}$, with a mass and length
scale $u_0$ in the problem.

\bibitem{RaB98}
M.A. Ramos and U. Buchenau, in: {\em Tunneling Systems in  Amorphous and
Crystalline Solids}, P. Esquinazi (Editor), (Springer, Berlin 1998),  p. 527

\bibitem{HoKu99}
The decrease at larger $T$ is due to the fact that only one phonon
processes  are considered as in U. Horstmann and R. K\"uhn, Physica B {\bf
263--264}, 290 (1999)

\end{thebibliography}
\end{document}